\documentclass[twocolumn,showpacs,preprintnumbers,prb]{revtex4-1}

\usepackage{graphicx}
\usepackage{dcolumn}
\usepackage{bm}
\usepackage{amsmath}
\usepackage{amssymb}
\usepackage{hyperref}

\newcommand{\bB}{\mathbf{B}}

\newcommand{\bG}{\mathbf{G}}

\newcommand{\br}{\mathbf{r}}
\newcommand{\bR}{\mathbf{R}}

\newcommand{\bv}{\mathbf{v}}

\renewcommand{\vec}[1]{{{\mathbf{\boldsymbol #1}}}}
\newcommand{\cD}{\mathcal{D}}

\newcommand{\he}{\hat{e}}
\newcommand{\hn}{\hat{n}}

\begin{document}
\title{Magnetic Skyrmions on a Two-Lane Racetrack}
\author{Jan M\"uller}
\email{jmueller@thp.uni-koeln.de}
\affiliation{Institut f\"ur Theoretische Physik, Universit\"at zu K\"oln, D-50937 Cologne, Germany}
\date{\today}

\begin{abstract}
Magnetic skyrmions are particle-like textures in the magnetization, characterized by a topological winding number. 
Nanometer-scale skyrmions have been observed at room temperature in magnetic multilayer structures. 
The combination of small size, topological quantization, and their efficient electric manipulation makes them interesting candidates for information carriers in high-performance memory devices.  
A skyrmion racetrack memory has been suggested where information is encoded in the distance between skyrmions moving in a one-dimensional nanostrip. 
Here, we propose an alternative design where skyrmions move in two (or more) parallel lanes and the information is stored in the lane number of each skyrmion. 
Such a multilane track can be constructed by controlling the height profile of the nanostrip. 
Repulsive skyrmion-skyrmion interactions in narrow nanostrips guarantee that skyrmions on different lanes cannot pass each other. 
Current pulses can be used to induce a lane change. 
Combining these elements provides a robust, efficient design of skyrmion-based storage devices.
\end{abstract}

\maketitle

Magnetic storage devices today predominantly use the orientation of magnetic domains to encode huge amounts of information\cite{lit:MagneticStorageToday}.
The information density is limited by the size of the domains, which not only have to be thermally stable, but should also support features as easy and non-mechanically controlled writing and reading of information.
Magnetic skyrmions are particle-like textures in the magnetization\cite{lit:muehlbauer,lit:SkyrmionLatticeRealSpaceObservation,lit:SkyrmionLatticeAtomicScale} of nanometer size, which can be controlled by ultra-low electronic\cite{lit:SkyrmionMotionUltralowCurrent,lit:SkyrmionFlowUltralowCurrent,lit:SkyrmionMotionUniversalCurrentVelocity} or magnonic\cite{lit:SkyrmionLatticeSpinTorquesRotation,lit:SkyrmionMagnonScattering} current densities. 
Due to these properties, they are often treated as promising candidates for information carriers in high-density, non-volatile, solid state storage devices\cite{lit:SkyrmionStorageKiselev,lit:SkyrmionRacetrack}.

Since the experimental discovery of the skyrmion lattice in the chiral magnet MnSi at low temperatures\cite{lit:muehlbauer}, skyrmion lattices and also single skyrmions have been observed in many different systems.
The energetic stability of skyrmions can be explained by Dzyaloshinskii-Moriya interaction (DMI), a spin-orbit coupling effect which arises either from broken inversion symmetry within the unit cell\cite{lit:bogdanov,lit:SkyrmionLatticeGroundState,lit:SkyrmionLatticeInTwoDimensions} (bulk DMI) or inversion broken by an interface\cite{lit:SurfaceDMI} (interfacial DMI).
In the altter case the thermal stability of skyrmions could be enhanced by a chiral stacking of thin films\cite{lit:InterfacialDMIstacking}, so that skyrmions have recently been stabilized at room temperature in these multilayer systems\cite{lit:SkyrmionsAtRoomTemperature}.

\begin{figure}[t]
  \centering
  \begin{minipage}[b]{0.47\textwidth}
    \centering
    \includegraphics[width=0.95\textwidth]{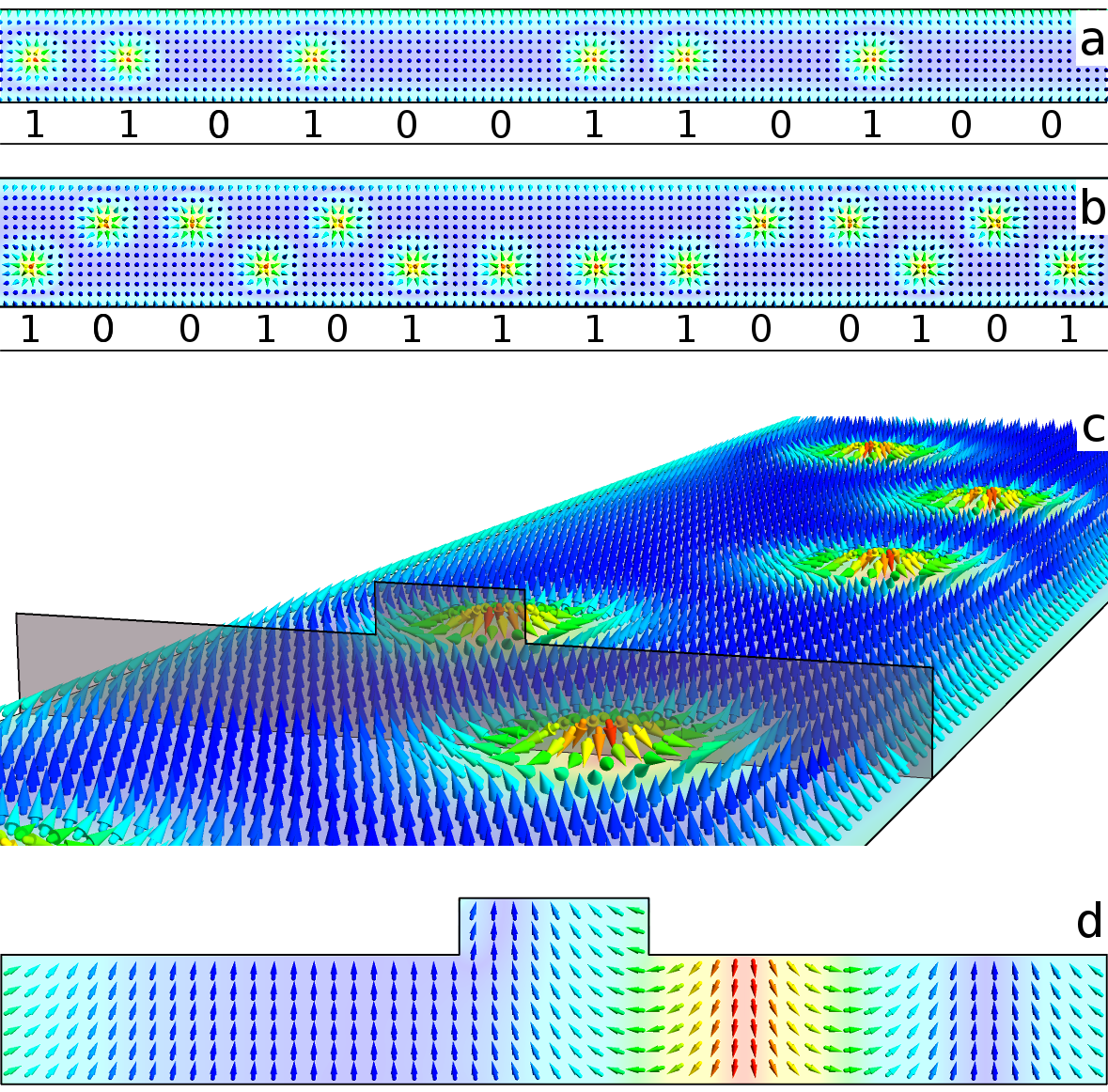}
    \caption{{\bf Skyrmion racetrack models.}
	     The color code denotes the z-component of the magnetization.
	     Only a subset of the simulated spins is shown. 
	     (a) One-lane skyrmion racetrack as proposed by Fert {\it et al}\cite{lit:SkyrmionRacetrack}: 
	     The information is encoded in the skyrmion distance.
	     (b) Two-lane skyrmion racetrack (top view of bottom layer): 
	     Here the information is encoded in the lane index of each skyrmion.
	     (c) Magnetization in the bottom layer and profile (gray area) of the two-lane racetrack. 
	     The two lanes are separated by a region of increased height.
	     (d) Magnetization in the gray shaded area of (c).
	     }
    \label{fig1}
  \end{minipage}
\end{figure}

Experimentally, a wide range of techniques is now used to image single skyrmions, e.g. Lorentz transmission electron microscopy\cite{lit:SkyrmionLatticeRealSpaceObservation}, magnetic force microscopy\cite{lit:monopoles}, spin-polarized scanning tunnelling microscopy\cite{lit:SkyrmionLatticeAtomicScale} or also X-ray based techniques, e.g. magnetic transmission soft X-ray microscopy\cite{lit:SkyrmionCurrentDrivingExperiment}.
For applications in memory devices it is important that single skyrmions can also be detected purely electronically exploiting their non-coplanar magnetoresistance\cite{lit:SkyrmionElectronicDetection} or the topological Hall effect\cite{lit:SkyrmionLatticeTopologicalHallEffect}.

On the road towards applications, controlled writing and deleting of single skyrmions has been shown to work in experiments using current injection\cite{lit:SkyrmionWriteAndDeleteRomming}.
The creation of chains of skyrmions at the edge of a sample has also been demonstrated experimentally\cite{lit:SkyrmionChainExp}.
Other processes for the creation of single skyrmions near the edge have been proposed theoretically\cite{lit:SkyrmionWriteAndDeleteKoshibae,lit:EdgeInstability}.
Also the driving of a chain of skyrmions by electric current in a nanowire was shown in experiments\cite{lit:SkyrmionCurrentDrivingExperiment}.

A prominent model for such a skyrmion-based storage device is the skyrmion racetrack\cite{lit:SkyrmionRacetrack}, a nanostrip with a one-dimensional distribution of skyrmions, Fig.~\ref{fig1}a.
The information in this device is encoded in the distance between the individual skyrmions.
By applying an electric current along the track, the train of skyrmions is pushed along the nanowire towards fixed read and write elements of the memory.
Since the drift velocity of the skyrmions is increased by the Magnus force when they are pushed against the edges of the track, the translation can be designed more efficient by using perpendicular driving forces, e.g. by the spin Hall effect\cite{lit:SkyrmionNucleationStabilityMotion}.
This way, a non-volatile, non-mechanic, and high-density magnetic memory device is created\cite{lit:SkyrmionRacetrack}.

As the information in this device is encoded in the distance between skyrmions, this distance needs to be preserved on the operational time scales.
Here the model has to fight problems:
Thermal diffusion\cite{lit:SkyrmionInertiaDiffusionDynamicsSchuette} and, more importantly, the noise created when operating the memory devices at high frequencies in a disordered environment will lead to a redistribution of the skyrmion distances. 
Furthermore the interaction of skyrmions is repulsive\cite{lit:SkyrmionSkyrmionEdgeRepulsion}.
These problems can however be solved, if the continuous translational invariance along the track is reduced to a discrete one.
Devices based on the racetrack have been proposed, that for example use a regular arrangement of notches\cite{lit:SkyrmionRacetrackNotchedAndGated} or electric potentials\cite{lit:SkyrmionRacetrackVoltageControlled,lit:SkyrmionRacetrackComplementary} to divide the track into a sequence of parking lots for skyrmions.
Such obstacles can, however, hinder the motion of the skyrmions towards read or write elements. 
Due to the complex pulsed motion of the skyrmions also disorder effects may become more important\cite{lit:SkyrmionHole,lit:SkyrmionMotionAndPinningByDefectsExperiment}.

In this paper we present a new ansatz for the application of skyrmions as information carriers in racetrack layout devices, which combines digital encoding of information with the benefits of continuous driving.
We prepare a racetrack with not one but two lanes on the same strip, Fig.~\ref{fig1}b, which are separated by a high enough energy barrier such that skyrmions do not change the lane by thermal activation.
The barrier can have various origins, but in this paper we focus on an additional nanostrip on top of the racetrack.
This creates a repulsive potential, Fig.~\ref{fig1}c, complementary to the idea in Ref.~\citenum{lit:SkyrmionRacetrackPotentialWell} where an attractive potential is created from a notch on the racetrack.
The two lanes are chosen to be sufficiently close that skyrmions on different lanes repulsively interact and therefore can not pass each other.
The main difference to the original skyrmion racetrack concept is the encryption of information:
It is encoded in the index of the occupied lane.
When applying an electric current to this densely packed sequence of skyrmions, they can still move along the track with all the speed-up benefits of the original skyrmion racetrack\cite{lit:SkyrmionNucleationStabilityMotion}.
A writing element is installed on a fixed position, where it induces a lane change, for example by applying a strong electric current pulse.
With the redefinition of the logic bit, thermal diffusion does not play a role anymore as long as the potentials are large enough.
Also perturbations by disorder only become relevant if they are of the order of the artificial potentials between skyrmions and the nanostructure or other skyrmions, respectively.

\section{Potentials and landscapes}
\label{sec:potentials}
\begin{figure}[t!]
  \centering
  \begin{minipage}[b]{0.47\textwidth}
    \centering
    \includegraphics[width=0.95\textwidth]{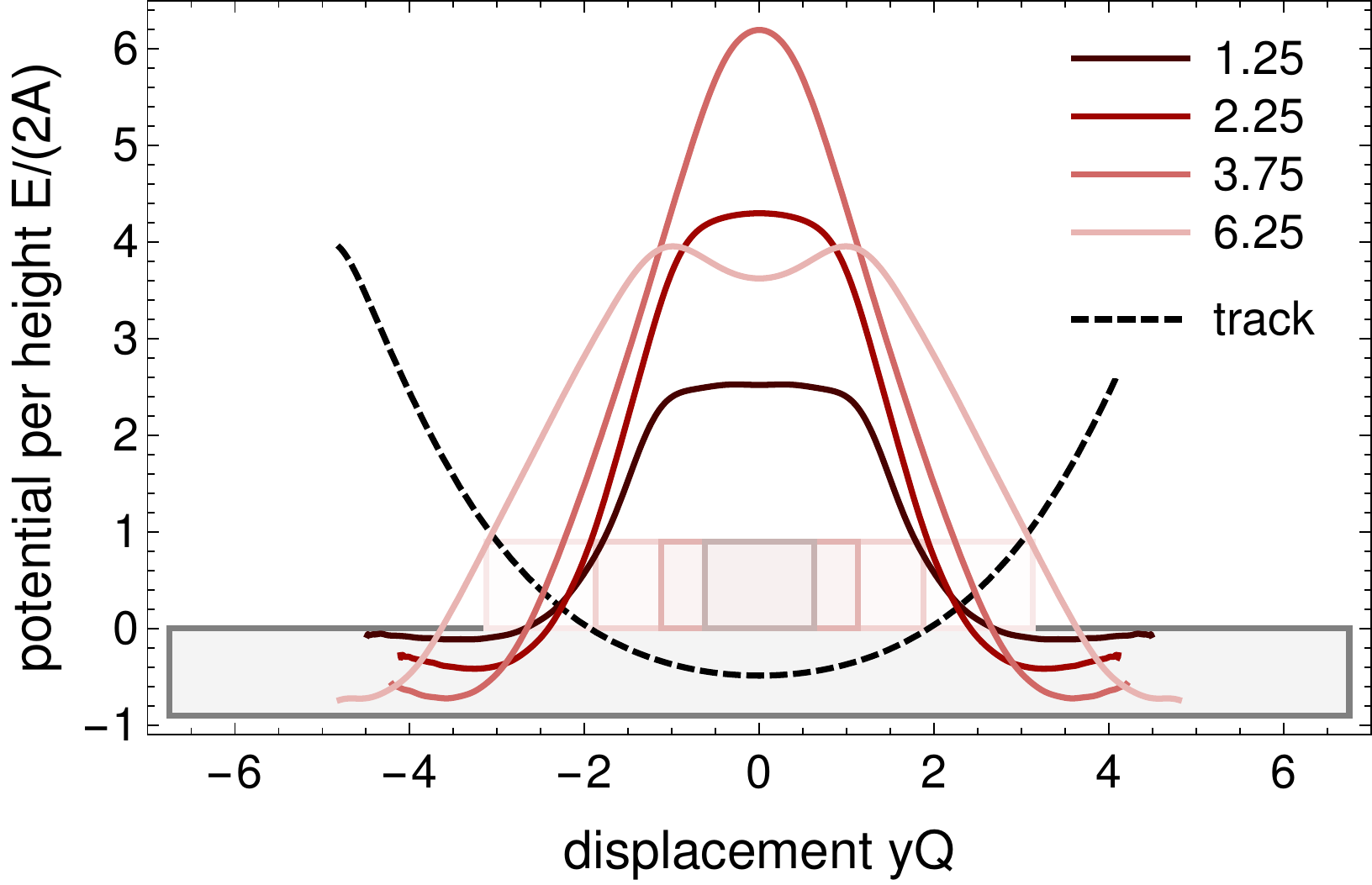}
    \includegraphics[width=0.95\textwidth]{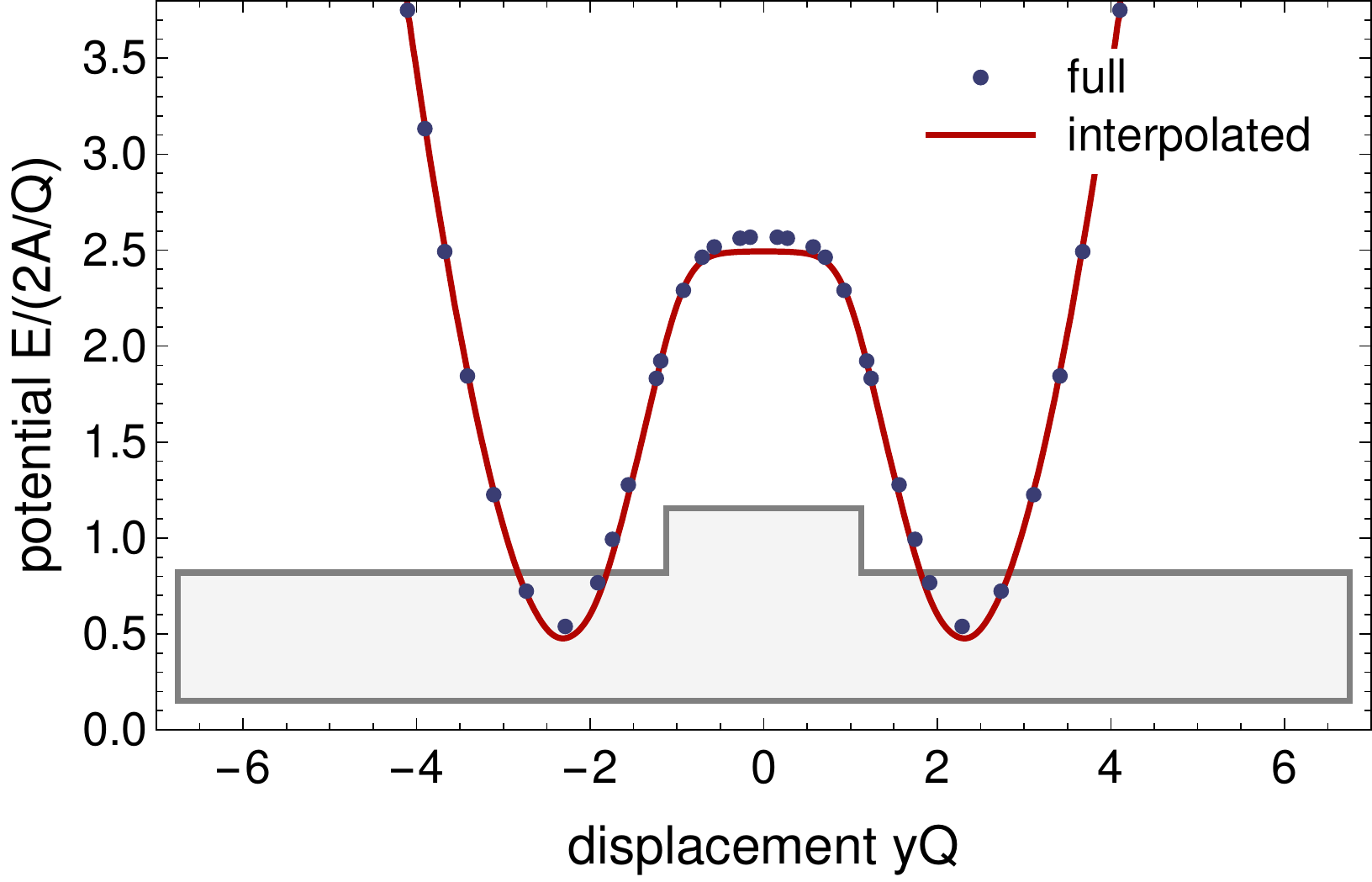}
    \includegraphics[width=0.95\textwidth]{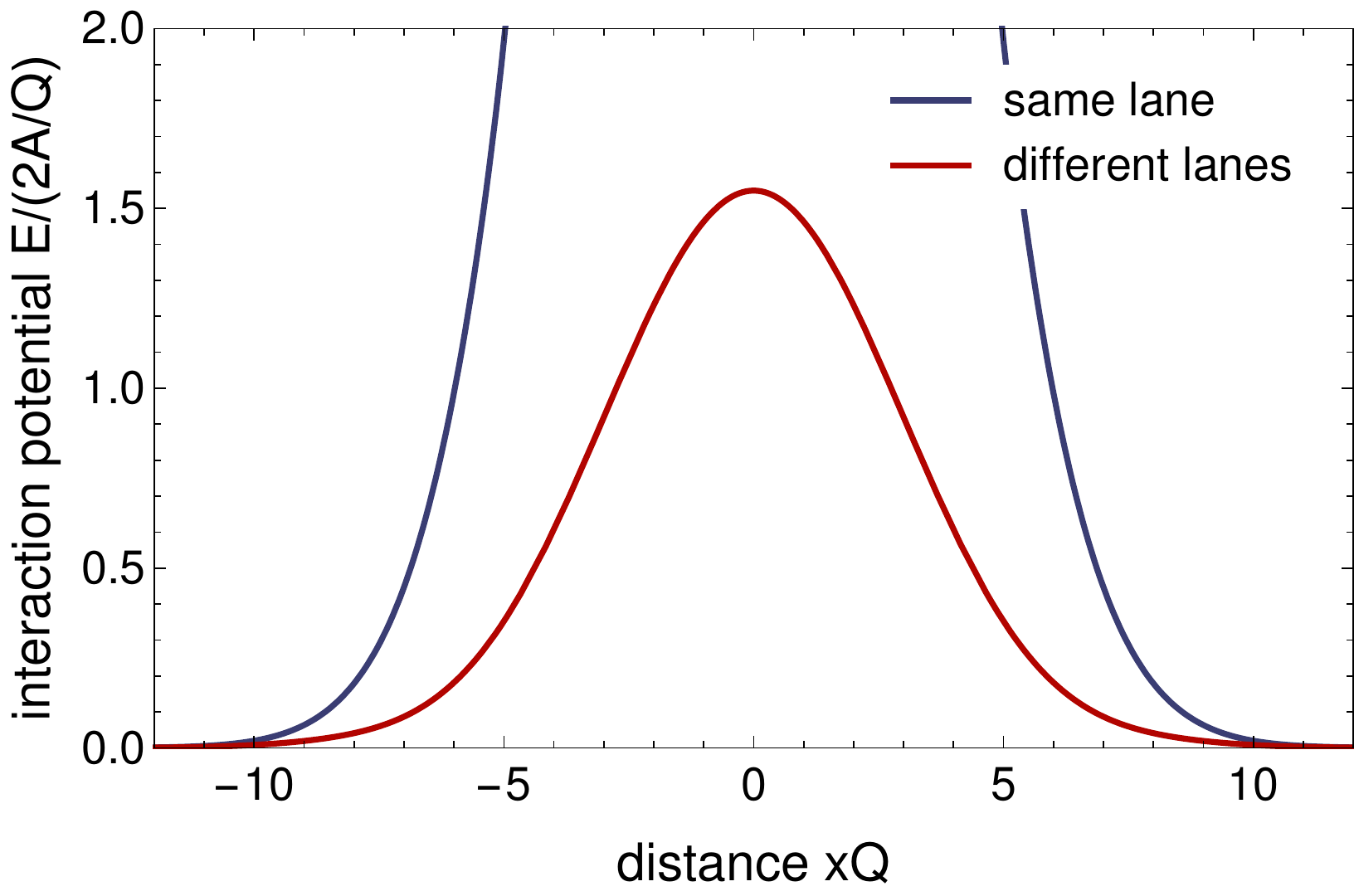}
    \caption{{\bf Skyrmion potentials in a nanostructured racetrack}.
	     The effective potential of the skyrmion in the nanostructure can be obtained by adding potentials for the bottom layers and the barrier in the center scaled by their respective heights. 
	     All figures show results for a magnetic field $H=0.75 h_D$.
	     Upper panel: Potential per height for the bottom layers (dashed line) and various central barriers (solid lines).
	     Bottom layer width is $13.75/Q$ (gray shaded area), presented barrier widths are $1.25/Q$, $2.25/Q$, $3.75/Q$, and $6.25/Q$ (colored shaded areas).
	     Middle panel: Comparison of the full skyrmion potential (dots) to the interpolated potential (line) for a nanostructure with bottom height $1.50/Q$, barrier height $0.75/Q$, barrier width $2.25/Q$ (gray shaded area).
	     Lower panel: Interaction potential of two skyrmions on the same lane (blue) and distinct lanes (red) as a function of distance along the track.}
    \label{fig2}
  \end{minipage}
\end{figure}

The operation of a two-lane racetrack builds on 
(i) a potential which keeps the skyrmions on the lanes, 
(ii) the skyrmion-skyrmion repulsion both for skyrmions on the same and on different lanes and 
(iii) a mechanism to let a skyrmion change lanes.
We calculate the racetrack and interaction potentials based on a micromagnetic model which considers the exchange coupling $A$ of spins, interfacial DM interactions $D$ and an external magnetic field $H$, see methods~\ref{sec:methods}.
We have checked that a realistic anisotropy does not change the main results qualitatively and thus neglect anisotropies and also dipolar interactions for simplicity.
Within our setup, all results depend only on a small number of dimensionless parameters, see methods~\ref{sec:methods}.
Most imporantly, all length scales are measured in units of $1/Q=2A/D$, which ranges from 1-100 nm for different types of skyrmion realizations\cite{lit:SkyrmionMotionAndPinningByDefectsExperiment,lit:SkyrmionCurrentDrivingExperiment,lit:SkyrmionsAtRoomTemperature}.
The field $H=0.75 h_D$ is chosen such that it stabilizes single skyrmions and they do not decay into bimerons\cite{lit:Bimerons}.

The effective skyrmion potential can be designed by changing the height and width of the track and the barrier. 
For fixed widths, we find that for a wide range of parameters, the dependence on the height of the track and the barrier can to high accuracy be obtained from a simple linear relation
\begin{equation}
V(y) = h_{\text{track}} v^\text{track}(y) + h_{\text{barrier}} v^\text{barrier}(y) \text{.}
\label{eq:potential_skyrmion-single_superposition}
\end{equation}
The simple additivity reflects that deformations of the magnetic texture along the track normal are small and it may be useful to design track potentials in experimental systems.
Qualitatively the shape of the potentials can be understood from the fact that skyrmions are repelled by edges\cite{lit:SkyrmionSkyrmionEdgeRepulsion}.
Therefore not only the track potential raises towards the edge of the track but also the barrier itself is repulsive, see Fig.~\ref{fig2}a.
The magnetic texture is continued from the bottom layers into the nanostructured upper layers, where effectively the skyrmion is in close proximity to an edge.
We find the largest potential barriers if the strip and the skyrmion are of comparable width, compare Fig.~\ref{fig2}a.
If the strip is very broad, the skyrmion in the center of the track is again relatively far away from the edges in the barrier structure.
The exact shape of the potential, however, strongly depends on the width of the strip.

In the following, we will focus on a particular example of a two-lane racetrack.
We consider a track with a width of $13.75/Q$ and a height of $1.50/Q$.
The additional nanostructure on top is $2.25/Q$ in width and $0.75/Q$ in height, see Fig.\ref{fig1}d.
From the scaling argument, Eq.~\ref{eq:potential_skyrmion-single_superposition}, an estimation of the potential yields a double-well shape with two degenerate minima.
As a check, we calculated the full potential of this geometry and find that the results are in very good argeement, see Fig.~\ref{fig2}b.

Finally, we also calculated the potential of two interacting skyrmions where we have to distinguish the two cases where skyrmions move in the same or in different lanes, see methods~\ref{sec:methods}.
In both cases the interactions are always repulsive, see Fig.~\ref{fig2}c, and scale linearly in the height of the structure.
Note that for the chosen parameters the repulsive interaction of skyrmions in different lanes is sizable and comparable to the height of the barrier between the lanes.

\section{Operation of a two-lane racetrack}
\label{sec:operation}

The above analysis of the separating potential provides a toolbox to engineer the barrier in a way that the order of the skyrmions is conserved with the help of repulsive potentials.
For the operation of the two-lane racetrack as a memory device, a read and write protocol is needed.

\begin{figure}[t!]
  \centering
  \begin{minipage}[b]{0.47\textwidth}
    \centering
    \includegraphics[width=0.95\textwidth]{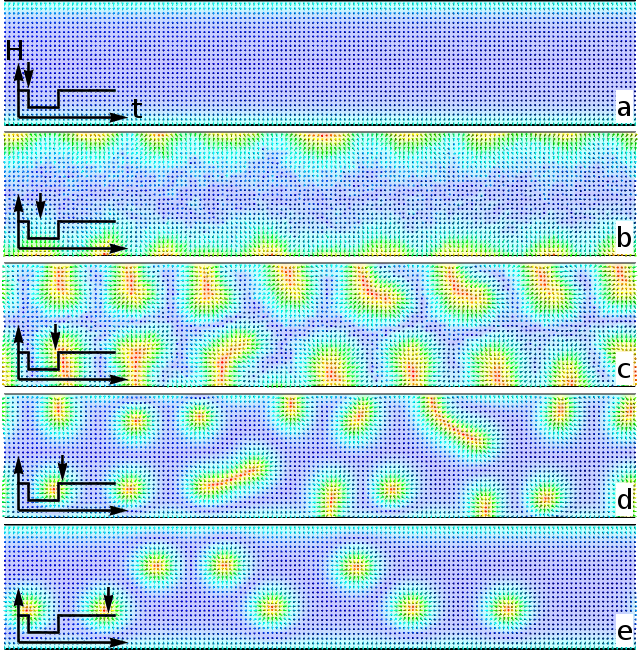}
    \caption{{\bf Creation of skyrmions.} 
	     Topview on the bottom layer at times $t/t_D=0, 30, 60, 80, 200$.
	     We start from the polarized ground state with an external magnetic field $H=0.75\,h_D$ (a).
	     The field is suddenly reduced to $H=0.3\,h_D$ (b) (plus small fluctuations) which triggers the edge instability:
	     Magnons condense at the edges (b) and merons enter (c).
	     Upon suddenly increasing the field again to $H=0.75\,h_D$, the merons either leave the system or pull a second meron from the edge to turn into a skyrmion (d).
	     The skyrmions redistribute over the two-lane track (e).
	     See also Supplementary Movie 1.
	     }
    \label{fig3}
  \end{minipage}
\end{figure}

The thermodynamic ground state of the two-lane racetrack is the polarized phase for parameters discussed above. 
Due to their topology, metastable skyrmions are, however, even at room temperatures very stable \cite{lit:SkyrmionCurrentDrivingExperiment}.
To initialize the track, i.e. fill it with skyrmions, we suggest to simply lower the magnetic field for a short period, see Fig.~\ref{fig3} and Supplementary Movie 1.
As discussed in Ref.~\citenum{lit:EdgeInstability}, this triggers an edge instability where magnons with finite momentum condense at both edges.
The polarized phase then becomes unstable and a chain of merons invades from the edges.
Technically, one has to add a small noise term or an edge inhomogeneity to trigger the magnon condensation.
In an experimental realization this is done by defects and thermal fluctuations. 
By afterwards increasing the field above the critical field to its original value, each meron either pulls a second meron from the edge to form a skyrmion, or is pushed back into the edge and vanishes.
Thereby the system automatically generates a finite density of skyrmions.
We find that the average distance between skyrmions, here approximately $8/Q$, is equal to the wavelength of the edge instability.
The initialized configuration, however, is random.
Lowering the external field on only one lane generates a homogeneous initialization but is technically more difficult.

For a large operational speed and maximal durability, the read and write elements should be non-mechanic.
We suggest to install them both in a combined operational unit as a write operation requires reading the state first.
Reading information can be realized in various ways.
One option is to use a tunnel contact on top or below each lane which exploits the non-collinear magnetoresistance of a skyrmion\cite{lit:SkyrmionElectronicDetection}.
Alternatively, one could also use the topological Hall effect\cite{lit:SkyrmionLatticeTopologicalHallEffect} to detect skyrmions.

\begin{figure}[t!]
  \centering
  \begin{minipage}[b]{0.47\textwidth}
    \centering
    \includegraphics[width=0.95\textwidth]{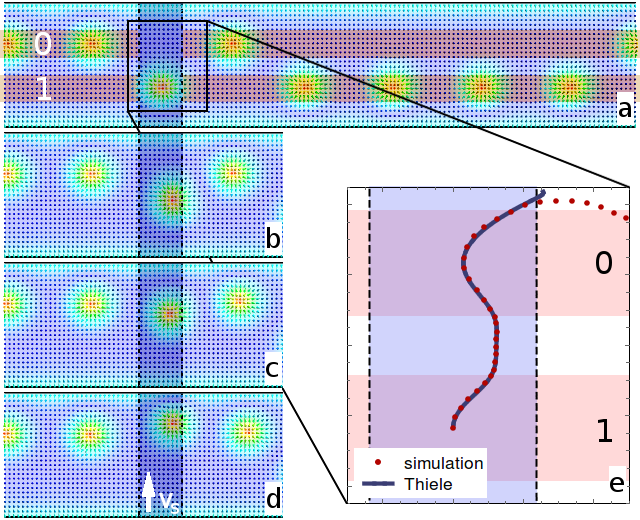}
    \caption{{\bf Writing information}.
	     An electric current $j$ flows in the blue shaded area ($\bv_s=0.1 v_D \,\he_y$).
	     Initial state (a) and snapshots (b-d) in time steps $\Delta t=25\,t_D$: Lane change of a skyrmion under applied current.
	     See also Supplementary Movie 2.
	     Comparison (e) of the skyrmion trajectories obtained from a simplified Thiele model (blue) and the micromagnetic simulation (red).
	     }
    \label{fig4}
  \end{minipage}
\end{figure}

Once a skyrmion is detected, one can decide if it has to switch lanes or already represents the desired information.
The lane change, in general, can be induced by a variety of forces.
While in insulating materials magnonic currents provide an extremely efficient method to control skyrmion motion in nanostructures\cite{lit:SkyrmionRacetrackDrivingMagnons}, we restrict our considerations to only electric currents.
The here considered write element consist of a pair of nanocontacts attached to both edges of the track, such that an applied electric current pulse flows predominantly through a narrow stripe as depicted in Fig.~\ref{fig4}.
The stripe has to be narrow such that only one skyrmion is affected by the current pulse.
For our calculations, we choose a stripe with a width of $5/Q$.
The minimal current density required to induce a lane change can be roughly estimated from the separating potential barrier $V(y)$ in a Thiele approach\cite{lit:Thiele}.
The simplified Thiele equation, the equation of motion for the skyrmion coordinate $\bR$, see methods~\ref{sec:methods}, reads:
\begin{align}
-\frac{1}{h_\text{track}} \frac{d V}{d \bR} =
 \mathbf{G} \times \left( \dot{\bR} -  \bv_s \right)
 + \mathcal{D}  \left( \alpha \dot{\bR} - \beta \bv_s \right) \text{.}
\end{align} 
A suitable current density for a lane change $v_s$ requires that during the lane change, the transversal shift of the skyrmion is smaller than the width of the current carrying region.
Solving the Thiele equation numerically for various $v_s$, which can easily be done, we find that these requirements are fulfilled for $v_s=0.1\,v_D$. 
Comparing the simplified Thiele estimate to the full micromagnetic simulation, we find that they are in very good agreement, see Fig.~\ref{fig4}, and the current density is indeed suitable for a writing process.
In the simulation, we applied the current pulse significantly longer than necessary for a lane change, to explore the skyrmion motion beyond the limits of our simple Thiele approach.
Instead of leaving the track, the skyrmion pushes its neighbor (which is artificially fixed in the Thiele ansatz) and all following skyrmion in the train to the side, while the order is still preserved.
Consequently, for a writing process a pulse time of $\Delta t = 50-75\,t_D$ is sufficient, see Figs.~\ref{fig4}(a-d), but the writing operation is very robust against disturbance in the process timing.

Since the write and read elements are fixed in position, the skyrmions have to be moved.
The current-driven motion of skyrmions in a nanostructure has been nicely discussed in Refs.~\citenum{lit:SkyrmionMotionConstrictedGeometries} and \citenum{lit:SkyrmionNucleationStabilityMotion}.
As previously pointed by the authors, one can obtain much higher velocities for skyrmions moving in racetracks compared to freely moving skyrmions.
Alternatively, the skyrmions can be driven highly efficient by a magnonic current\cite{lit:SkyrmionMagnonScattering,lit:SkyrmionRacetrackDrivingMagnons}.
Using one of these methods, the bit sequence is driven along the two lanes, see also Supplementary Movie 2.

\section{Summary}
We propose a new approach towards skyrmion-based storage devices.
The two-lane skyrmion racetrack uses the displacement of skyrmions in a racetrack relative to the center of the track to encode information.
We provide methods to calculate the potential barrier between different lanes as well as the interaction between skyrmions in the lanes.
In this paper we show that these potentials can be used to efficiently design such a two-lane racetrack and to determine current pulses to write information.
With these informations we explain how to operate a particular example of a two-lane racetrack by electric current pulses and confirm the results by micromagnetic simulations.
Dispite that the model we used throughout this work does not take into account uniaxial anisotropy or dipolar interactions, which may be of interest for realistic systems, the idea presented here is universal.
That the idea of a two-lane racetrack combines the powerful idea of a skyrmion racetrack with the promising properties of the skyrmion lattices, i.e. low depinning current thresholds and highest information carrier densities whilst the information is protected against perturbations.
We believe that designs based on the two-lane skyrmion racetrack are promising candiates for a future skyrmion-based information technology.

\section{Methods}
\label{sec:methods}
\subsection{Model and units}
The starting point of our analysis is a non-linear sigma model in which the magnetization is described by the normalized vector field $\hn^T = (\hn_x, \hn_y, \hn_z)$ with $\|\hn\|=1$.
The free energy functional, $F = \int d^3 \vec r \mathcal{F}$, with the minimal set of interactions that we choose in this work, contains only the ferromagnetic exchange interaction $A$, interfacial Dzyaloshinskii-Moriya interaction $D$ and an external magnetic field $H$:
\begin{align}
\mathcal{F} = A (\partial_\alpha \hat{n}_\beta)^2 + D \left( \hn_\gamma \partial_\gamma \hn_z-\hn_z \partial_\gamma \hn_\gamma   \right) - \mu_0 H M \hat n_z \text{,}
\label{eq:energy}
\end{align}
with $\alpha,\beta = x,y,z$, $\gamma = x,y$ and $M$ the saturation magnetization.
The scales for momentum, total energy, and magnetic field are chosen in accordeance with previous works
\begin{align} 
Q = \frac{D}{2A}, \quad  E_D = \frac{2 A}{Q}, \quad h_D = \frac{2 A Q^2}{\mu_0 M}
\end{align}
and are used throughout the work.
From the LLG equation 
\begin{eqnarray}
\left[\partial_\text{t} \!+\! \left( \mathbf{v}_s \!\cdot\! \nabla \right)\right]\! \hn= &-&\gamma \hn \times \mathbf{B_\text{eff}} \nonumber\\
&+& \alpha \hn \!\times\! \left[ \partial_\text{t} \hn + \frac{\beta}{\alpha} \!\left( \mathbf{v}_s \!\cdot\! \nabla \right) \hn \right] \text{,}
\label{eq:LLG}
\end{eqnarray}
with the effective magnetic field $\vec B_{\rm eff} = - \frac{1}{M} \delta F/\delta \hat n$ we deduce the dependence of the drift velocity of the spin currents $\mathbf{v_s}$ and the time $t$ on the spin denity $s=\frac{M}{\gamma}$:
\begin{align} 
t_D = \frac{s}{2AQ^2}, \quad  v_D = \frac{2AQ}{s} \text{.}
\end{align}
Note that $s$ is the spin per unit cell of the atomic lattice, i.e. independent of $Q$.

With realistic values\cite{lit:InterfacialDMIstacking} for the exchange coupling $A=10 \mathrm{\,pJ\,m^{-1}}$, interfacial DMI $D=1.9 \mathrm{\,mJ\,m^{-2}}$ and a saturation magnetization 
of $M=956\mathrm{\,kA /m}$, the corresponding magnetic field that we apply in our simulations is $\mu_0 H = 0.75 \frac{2 A Q^2}{M} = 0.14\mathrm{\,T}$.
Note that in systems with uniaxial anisotropy skyrmions of similar size can be realized with considerably smaller (or even vanishing) external magnetic field.
Here, the unit of length corresponds to approximately $1/Q = 10.5\,\mathrm{nm}$.
Hence our simulated racetrack has a width of $13.75/Q = 145\,\mathrm{nm}$ and a height of $1.5/Q = 16\,\mathrm{nm}$, while the strip is of $2.25/Q = 24\,\mathrm{nm}$ width and $0.75/Q = 8\,\mathrm{nm}$ height.
With the unit of energy $E_D = 2 A/Q= 2.1\times10^{-19} \,\mathrm{J}$ we obtain an activation energy for the barrier between the lanes of about $2 E_D/\mathrm{k_B} = 30,000 \,\mathrm{K}$ implying thermal stability at room temperature.
Finally, with the unit of time $t_D = \frac{s}{2AQ^2} = 0.03\,\mathrm{ns}$ (assuming $g=2$) the writing process with minimal current (of the order of $10^{10} \,\mathrm{A/m^2}$) as described in the main text takes only $1.5-2.2\,\mathrm{ns}$.

\subsection{Micromagnetic simulations}
For simulations of the continuous model, we discretize the magnetization $\hn$ to Heisenberg spins on a cubic lattice.
We approximate the derivatives in Eqs.~\ref{eq:energy} and \ref{eq:LLG} by finite differences.
The lattice constant is $a=0.25/Q$ throughout the paper in all three spatial directions.
It is small enough that the results are independent of the discretization length.
We have written a code which integrates the discretized LLG equation over time with a fourth order Runge-Kutta-method.
The boundary conditions are such that the wire is periodic in the extended direction and open in the other directions.
The initial states in Sec.~\ref{sec:potentials} are polarized along the direction of the magnetic field.
For the energies as a function of the skyrmion position, we manually inserted a single skyrmion and for the skyrmion interaction two skyrmions, respectively.
The initial state in Sec.~\ref{sec:operation} is polarized and skyrmions are created automatically as a result of the system dynamics.
In the creation process, we added a small random fluctuating field to $\bB_\text{eff}$, see Eq.~\ref{eq:LLG}.

\subsection{Calculation of the potentials}
We calculated the skyrmion potential $V(y)$ numerically as the energy difference between a skyrmion at a given displacement from the center of the track $y$ and the polarized state without skyrmions:
\begin{equation}
V(y) = E_\text{skyrmion}(y) - E_\text{polarized}
\label{eq:potential_skyrmion-single_definition}
\end{equation}
The position of a skyrmion is here defined as the coordinate in the bottom layer, where the magnetization points antiparallel to the external field $H$.
Previous works fixed this center of the skyrmion to evaluate the potential at various positions\cite{lit:SkyrmionParticleModelReichhardt,lit:SkyrmionHole}.
In this setup, however, the forces from the nanostructures are large compared to the artificial pinning forces.
Therefore we can not fix the skyrmion coordinate.
We resolve this problem by a dynamical evaluation of the potential, i.e. we record the energy while running a simulation of the Landau-Lifshitz-Gilbert equation in which the skyrmion moves adiabatically, i.e. slow enough that no internal modes are excited. 
In order to guide the skyrmion to different positions, we apply a sufficiently large current density, here $v_s=0.032 v_D$ and $\alpha=\beta=0.1$ for the potential of the track.
For these potentials of the skyrmion-skyrmion interaction, we define an inhomogeneous current density ($v_s=0.128 v_D$) which either drives the skyrmions into a collision if they start on the same lane or drives them in different directions if they start on different lanes, such that they pass each other.
Note that when running on different lanes, the magnus force breaks the mirror symmetry of the motion, i.e. it makes a difference if the skyrmions run on the right or the left lane.
For skyrmions on the same lane, the magnus force can induce a lane change if the damping is low ($\alpha\lesssim1$).
As we are interested in the low energy limit, we suppress these effects by a relatively large damping $\alpha=\beta=2$.

\subsection{Thiele analysis}
In the Thiele analysis one assumes that the dynamics of the magnetic texture reduces to an effective particle coordinate $\bR(t)$ in the two-dimensional plane, $\hn(\br-\bR(t)))$.
This assumption is clearly violated in the system, but as the skyrmion centers are to good approximation local and rigid objects, it still holds to large extents.
Along the track normal, we further assume that the magnetic texture is mainly constant, hence $\bR(t)$ is equal in all layers of total height $h_\text{track}$.
Thus the system is effectively two-dimensional.
The Thiele equation is then recovered by first multiplying the LLG equation from the left with $\hn\times$ and then projecting it down onto the translational mode by multiplying the equation with $\frac{d \hn}{d R_i}$ and integrating over space $\int d^3r$.
Finally, this results in the equation of motion for the skyrmion coordinate $\bR(t)$ in the two-dimensional track plane $(x,y)$:
\begin{align}
-\frac{1}{h_\text{track}} \frac{d V}{d \bR} =
 \mathbf{G} \times \left( \dot{\bR} -  \bv_s \right)
 + \mathcal{D}  \left( \alpha \dot{\bR} - \beta \bv_s \right)
\label{eq:simpleThiele}
\end{align} 
where only the potential $V(\bR)$ is obtained from a full three-dimensional calculation.
We approximated the values for $\bG$ and $\cD$ with the values for a single skyrmion in an infinite extended polarized background in two dimensions.
For the analysis of a lane change, we assume a constant current density in the whole system.
However we note that the Thiele ansatz then only holds as long as the skyrmion is at sufficient distance ($\gtrsim1/Q$) to the edges of the area where the current density is actually applied.

\section{Acknowledgements}
J.M. thanks A.~Rosch, V.~Cros, A.~Fert, and M.~Kl\"aui for the stimulating discussions.
This work was supported by Deutsche Telekom Stiftung and the Bonn-Cologne Graduate School of Physics and Astronomy BCGS.
We furthermore thank the Regional Computing Center of the University of Cologne (RRZK) for providing computing time on the DFG-funded High Performance Computing (HPC) system CHEOPS as well as support.

\section{Author contributions}
J.M. performed the numerical calculations, analyzed the data and wrote the manuscript.

\section{Competing financial interests}
J.M. has a German patent application related to this work, number $10\,\,2016\,\,200\,\,161.2$, 'Logischer Speicher mit einer Vielzahl von magnetischen Skyrmionen'.

\end{document}